\documentclass{aa}

\usepackage[varg]{txfonts}
\usepackage{natbib,graphicx}

\bibpunct{(}{)}{;}{a}{}{,} % to follow the A&A style

\begin{document}

\title{Master equation theory applied to the redistribution of polarized 
radiation in the weak radiation field limit}

\subtitle{VI. Application to the second solar spectrum of the \ion{Na}{i} D1 and D2 lines: convergence}

\author{V\'eronique Bommier}

\institute{LESIA, Observatoire de Paris, Universit\'e PSL, CNRS, Sorbonne Universit\'e, Universit\'e de Paris 
\newline 5, Place Jules Janssen, 92190 Meudon, France}

\titlerunning{Redistribution of polarized radiation VI. 
Application to the \ion{Na}{i} D1 \& D2 lines: convergence}
\authorrunning{V. Bommier}

\date{Received ... / Accepted ...}

\abstract
% context
{This paper presents a numerical application of a self-consistent theory of partial redistribution in nonlocal thermodynamical equilibrium (NLTE) conditions, developed in previous papers of the series.}
% aims
{The code was described in IV of this series. However, in that previous paper, the numerical results were unrealistic. The present paper presents an approximation able to restore the reliability of the outgoing polarization profiles.}
% methods
{The convergence of the results is also proved. It is demonstrated that the step increment decreases like $1/N^\alpha$, with $\alpha > 1$.}
% results
{Thanks to these additions, the results series behaves like a Riemann series, which is absolutely convergent. However, convergence is not fully reached in line wings within the allocated computing time. Development of efficient acceleration methods would be desirable for future work.}
% conclusions
{Agreement between the computed and observed linear polarization profiles remains qualitative only. The discrepancy is assigned to the plane parallel atmosphere model, which is insufficient to describe the chromosphere, where these lines are formed. As all the integrals are numerical in the code, it could probably be adapted to more realistic and higher dimensional model atmospheres. However, this is time consuming for lines with a hyperfine structure, as in the \ion{Na}{i} D lines. The net linear polarization observed in \ion{Na}{i} D$_1$ with the Z\"urich Imaging Polarimeter ZIMPOL mounted on the McMath-Pierce telescope at Kitt Peak is not confirmed by the present calculations and could be an artefact of instrumental polarization. The presence of instrumental polarization could be confirmed by the higher linear polarization degree observed by this instrument in the \ion{Na}{i} D$_2$ line center with respect to the present calculation result where the magnetic field is not accounted for. At this precise point, the Hanle effect acts as a depolarizing effect in the second solar spectrum. The observed linear polarization excess is found to be of the same order of magnitude in both line centers, namely 0.1\%, which is also comparable to the instrumental polarization compensation level of this experiment.}


\offprints{V. Bommier, \email{V.Bommier@obspm.fr}}

\maketitle

\section{Introduction}

This paper is the last in a series aimed to develop a theory capable of partial redistribution (PRD) and line profiles in the atomic density matrix formalism well adapted to describing the polarized atom in nonlocal thermodynamical
equilibrium (NLTE)  conditions. The aim of this formalism is to write down and solve the statistical equilibrium equations for the atomic density matrix elements, which include sublevel populations but also coherences (or phase relationship) between sublevels, responsible in particular for the Hanle effect. The atom is submitted to an incident radiation and to collisions, which both enter the transition rates between density matrix elements. Once these elements have been computed by solving the statistical equilibrium equations, the radiative transfer equation coefficients may be derived, which are absorption matrix and emissivity for the Stokes parameters \citep[see, e.g., Eq. (6.85) of][]{Landi-Landolfi-04}, and the radiative transfer equation may then be integrated along each line of sight in the medium in order to recompute the radiation incident upon the atom in an iterative scheme.

Partial
redistribution and line profiles are introduced by repelling step by step the Markov approximation in the atom--radiation interaction hamiltonian, and by synthesizing a practicable final equation from the contribution of each step. This enables relationships between incoming and outgoing photons as two-step coupled processes and beyond. The principles of the calculation are each described in \citet{Bommier-97a}. This latter publication also introduces line profiles in the equations. The profile appears as an infinite but convergent and well-known series as a function of the development, which may then be summed up, leading to the final equations, which are then nonperturbative. Although the calculation principles are more general, the final equations given in \citet{Bommier-97a} are for a two-level atom with an unpolarized lower level. \citet{Bommier-97b} incorporates the magnetic field effect, of arbitrary strength, and derives the partial redistribution matrix for polarized radiation in the presence of an arbitrary magnetic field.

The following paper of the series, \citet{Bommier-16a}, presents the multilevel equations able to describe the NLTE problem for a multilevel atom embedded in a magnetic field. The statistical equilibrium equations are given for the multilevel atom (multiterm as well), which have to be resolved. This enables lower-level atomic polarization for example. The advantage conferred by this process is that PRD is included in the basic formalism itself in a self-consistent manner, whereas, usually, complete redistribution (CRD) is first solved in a multilevel scheme, and PRD is then added by considering the levels pair by pair \citep{Uitenbroek-89,Uitenbroek-01}. In \citet{Bommier-97a, Bommier-16a}, extension from CRD to PRD is simply accounted for via a new fourth-order term that enters the emissivity and is comprised of a product of  three profiles able to include more than two levels. An interesting feature appears in this very general formalism: the atomic density matrix statistical equilibrium has to be resolved for each atomic velocity class, and not for an averaged atom. This  additionally and simultaneously   enables velocity redistribution to be accounted for.

As an application, the so-called two-term redistribution matrix, which is the redistribution matrix for an atom with fine and/or hyperfine structure responsible for a series of lines connecting a lower and an upper term, as in for instance, the \ion{Na}{i} D line pair, or the \ion{Mg}{ii} h and k lines. This is indeed a multilevel problem and the solution was enabled following \citet{Bommier-16a}. This two-term redistribution matrix is given in \citet{Bommier-17}, with a corrigendum in \citet[thanks to Ernest Alsina Ballester]{Bommier-18}.

An application code of the multilevel theory is presented in \citet{Bommier-16b},  and is devoted to computing the theoretical linear polarization profile (``second solar spectrum'') of the \ion{Na}{i} D lines observed close to the solar limb, where linear polarization is formed by radiative scattering. Observations of this profile are provided, for example, in \citet[Figs. 2 and 3]{Bommier-Molodij-02} and in \citet[Fig. 2]{Stenflo-Keller-97} and \citet[Fig. 1]{Stenflo-etal-00a}. The D$_2$ linear polarization profile interestingly displays a line center sensitive to the Hanle effect, and two far wings probably insensitive to the magnetic field and then able to serve as a local reference for the zero-field polarization. 

The D$_1$ line polarization profile is a subject of debate: a net linear polarization is visible in \citet[Fig. 2]{Stenflo-Keller-97} and \citet[Fig. 1]{Stenflo-etal-00a}, whereas there is no net linear polarization in \citet[Figs. 2, 3]{Bommier-Molodij-02}, or in \citet[Figs. 2 and 3]{TrujilloB-etal-01}, or in \citet{Gandorfer-00}. Observations by \citet[Fig. 19]{Malherbe-etal-07} also display a net linear polarization in \ion{Na}{i} D$_1$. However, its shape is different from that of \citet[Fig. 2]{Stenflo-Keller-97} and \citet[Fig. 1]{Stenflo-etal-00a}. This net linear polarization is surprising from a theoretical perspective, because the upper term of the D$_1$ line is $3^2P_{1/2}$, with $J=1/2$ leading to unpolarizable term and line, even if there is hyperfine structure as in \ion{Na}{i}. The hyperfine structure acts as a depolarizing mechanism on an already zero polarization, and therefore the net polarization should remain zero. However, this does not rule out a spectral shape of this polarization, and investigating this shape was one of the motivations of the present theoretical computation.

Because of the hyperfine structure, the modeling of the \ion{Na}{i} D line polarization is a fully multilevel problem requiring the formalism presented in \citet{Bommier-16a}. We developed this application in \citet{Bommier-16b}. Nevertheless, the obtained profiles were very far from agreeing with observations (see Fig. 5). As the numerical code and methods are fully described in \citet{Bommier-16b}, we do not repeat them here. The present paper is devoted to concluding on this subject by presenting significantly improved results after an approximation described in Sect. \ref{results}. The calculation convergence is detailed in Sect. \ref{convergence}, and Sect. \ref{discussion} is a concluding discussion about the remaining gap between the theoretical and observed profiles.

In our calculations, collisional broadening and collisional transitions are fully accounted for, from computations by \citet{Roueff-74} for line broadening and \citet{Kerkeni-Bommier-02} for transitions both due to collisions with neutral hydrogen atoms, and by applying the semi-classical method of \citet{SahalB-69a,SahalB-69b} for collisions with electrons and protons. In particular, collisional transitions between the upper $J=1/2$ and $J=3/2$ are fully accounted for, which remains impossible in the two-term redistribution matrix approach, as discussed in \citet{Bommier-17}.

\section{Results}
\label{results}

\begin{figure*}
\resizebox{\hsize}{!}{\includegraphics{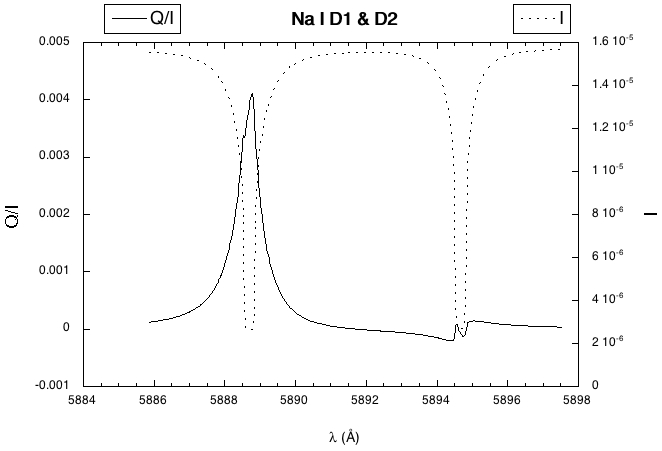}}
\resizebox{\hsize}{!}{\includegraphics{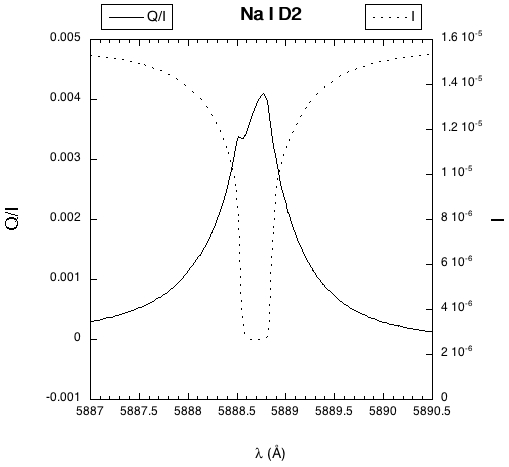}\includegraphics{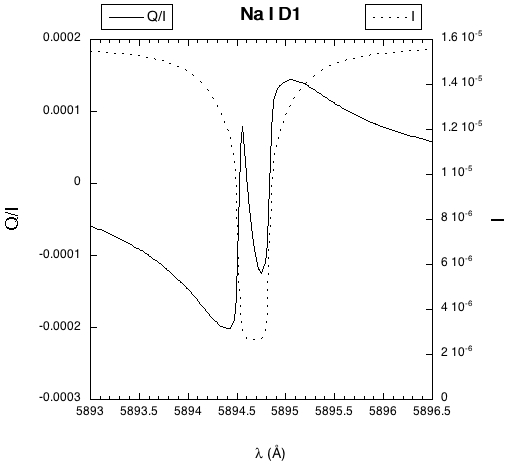}}
\caption{Intensity ($I$) and linear polarization ($Q/I$) theoretical spectrum of the \ion{Na}{i} D lines as observed 4.1 arcsec inside the solar limb. CRD: Assuming complete redistribution of radiation (second-order calculation). PRD: Including partial redistribution of radiation (infinite-order calculation).}
\label{profile}
\end{figure*}

The new results we present in this paper were obtained after some error corrections (in summations). Above all, a numerical approximation was carried out which is able to free the results from the unacceptable dispersion shape they show in Fig. 5 of  \citet{Bommier-16b}. The dispersion shape is due to the imaginary part of the profiles, which enter equations as complex quantities as visible in Eq. (63) of \citet{Bommier-97a}. As integrals are all numerical in the code, we attributed these certainly false dispersion profiles to insufficient accuracy of the integrals. As increasing the number of points and weights was impractical due to the long computing time, we forced the convergence by setting the imaginary part of the final profile  in the emissivity and absorption coefficient  to zero. As this appeared to bring about no effect on the fourth-order contribution responsible for the PRD, we finally did not apply it to this term. This was efficiently applied to the second-order contributions.

The emerging linear polarization is plotted in Fig. \ref{profile} together with the line intensity. The limb distance is taken at 4.1 arcsec ($\mu = 0.092$), as in the observation by \citet[Figs. 2 and 3]{Bommier-Molodij-02}. As discussed in \citet{Bommier-16b}, the atmosphere model is limited to the temperature minimum, as in the HOLMUL model \citep{Holweger-Muller-74}. This avoids a central bump in the line center intensity profile, which results from the temperature rise in the chromosphere, as visible in Fig. 12 of \citet{Bommier-16b}. The \ion{Na}{i} D lines are formed in the low chromosphere. The line center bump does not exist in observations. With this HOLMUL approximation, already applied in \citet{Bommier-16b}, the computed intensity profile is in better agreement with observations.

The computed polarization profile is now in much better agreement with observations. However, the agreement remains qualitative. A small bump is visible in the blue wing of \ion{Na}{i} D$_2$, similar but much smaller than the far wings visible in observations.

The shape of the \ion{Na}{i} D$_1$ polarization profile is very similar to that of observations, at least that of \citet[Figs. 2 and 3]{Bommier-Molodij-02}. This complex profile is comprised of two parts: central sharp peaks and broad wings. The amplitude of the linear polarization at the sharp peaks  is significantly weaker than in the observations. However, the amplitude of the broad wings is  very similar to that of the observations. Interestingly, the computations by \citet[Fig. 4]{Belluzzi-TrujilloB-13} are of a similar very low polarization level for their sharp peaks. However, the broad wings of their theoretical profile are very different from those of our calculations and those of the observations. The computed \ion{Na}{i} D$_1$ polarization profile seems perfectly antisymmetrical and free from any net linear polarization, which is not surprising because the \ion{Na}{i} D$_1$ line is globally unpolarizable.

In Fig. \ref{profile} the linear polarization in \ion{Na}{i} D$_1$ may appear weaker than in observations. However, this may be assigned to a perspective effect due to the strength of the linear polarization peak in \ion{Na}{i} D$_2$, which is 0.4\% in our calculations. We note that this strength is rather variable among observations, being 0.5\% in \citet[Fig. 2]{Stenflo-Keller-97} and \citet[Fig. 1]{Stenflo-etal-00a}, of 0.35\% for a different date and place of observations by \citet[Fig. 1 at $\mu = 0.1$]{Stenflo-etal-00b}, and of 0.3\% in the observation by \citet[Figs. 2]{Bommier-Molodij-02}. The depolarizing effect of the magnetic field may play a role in this line center polarization sensitive to the Hanle effect. Our calculations do not include any magnetic field effect in a first step.

\section{Convergence}
\label{convergence}

\begin{figure*}
\resizebox{\hsize}{!}{\includegraphics{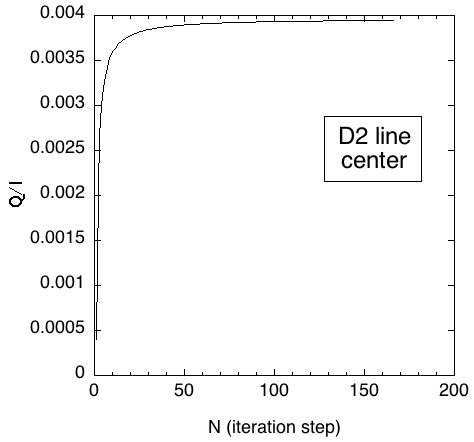}\includegraphics{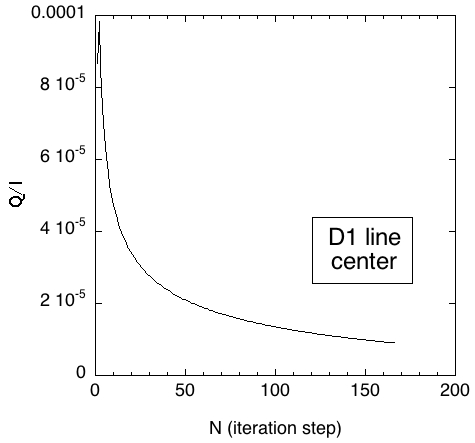}}
\resizebox{\hsize}{!}{\includegraphics{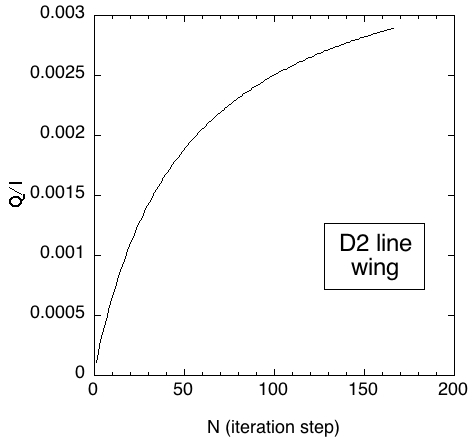}\includegraphics{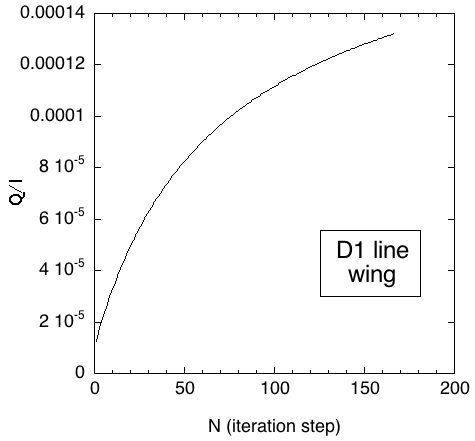}}
\caption{Convergence of the linear polarization rate $Q/I$, for the D$_2$ line (left) and for the D$_1$ line (right) at line center (upper row) and in the line far wing (lower row).}
\label{converge}
\end{figure*}

\begin{figure}
%\resizebox{\hsize}{!}{\includegraphics{bommierf3a}}
%\resizebox{\hsize}{!}{\includegraphics{bommierf3b}}
\includegraphics[width=9cm]{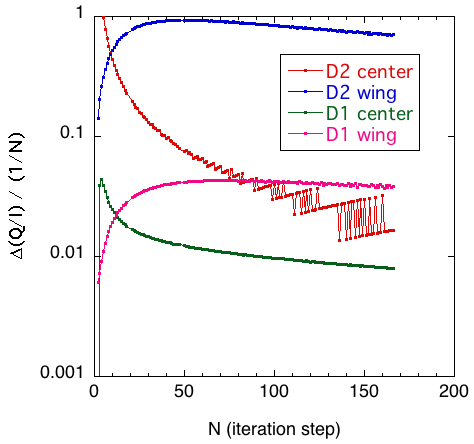}
\includegraphics[width=9cm]{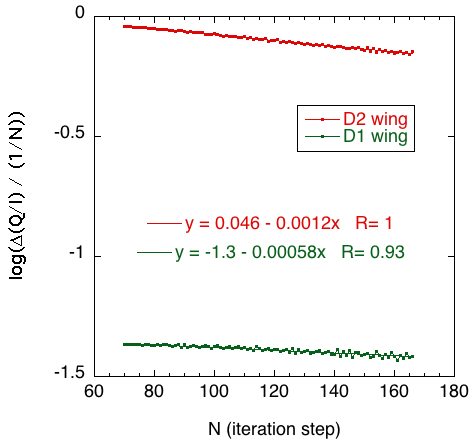}
\caption{Convergence of the emerging linear polarization $Q/I$: evolution of the step increment $\Delta (Q/I)$ as a function of the step number $N$. The aim is to show that the step increment decreases more rapidly than the inverse of the step number $1/N$ (as for a Riemann series). In this respect, their ratio is the upper plot and the logarithm of their ratio is the lower plot, which also includes linear fits and their equations.}
\label{riemann}
\end{figure}

This section is devoted to demonstrating that the results are essentially converged in terms of polarization degree $Q/I$. Indeed, the calculation method outlined at the beginning of this paper is of the lambda-iteration type, which is generally considered as poorly convergent. The implementation of acceleration methods was already discussed in \citet{Bommier-16b}. Below, we summarize and complement this discussion, and then present the results of the present calculations.

\subsection{Convergence acceleration}

Ng acceleration was attempted but was found inefficient, as reported in \citet{Bommier-16b}. Figure 3 of that paper shows that the emergent polarization is modified---sometimes increased, other times decreased---without any consistent convergence-improving trend. Indeed, this method is nothing more than a linear extrapolation, when the solution of the statistical equilibrium equations is linear in the two-level case only, which is not the present case.

The preconditioning would probably be inefficient for our density matrix comprised of 640 elements, in addition decomposed over 48 velocities. As demonstrated in \citet{SahalB-etal-98} and \citet{Bommier-16a}, the correct procedure for summing over atomic velocities is to solve the statistical equilibirum equations for each velocity class, and to integrate the emerging radiation over atomic velocities. The preconditioning would then be applied to a few density matrix elements of 1 velocity over 48. Weak efficiency is then expected. Alternatively, if different velocities become coupled by preconditioning, the full system  becomes $640 \times 48 = 30720$ elements in width, which would be huge. In addition, the complexity of the polarized atom formalism (Racah algebra), which is different for statistical equations \citep[see Eqs. (9-13) of][]{Bommier-16a} and radiative transfer equation coefficients \citep[see Eqs. (14-20) of][]{Bommier-16a} makes the preconditioning implementation such complex task that we renounced. We note that previous works successfully applied preconditioning up to ten levels only, without any velocity splitting up and with angle-averaged PRD for two by two levels \citep{Uitenbroek-01}. Given all these arguments, we decided not to implement any preconditioning, owing to its complexity in the polarized radiation and density matrix formalism, together with little hope of improvement because of the large number of matrix elements expanded over atomic velocities.

We studied the implementation of the Feautrier method in our formalism. The second-order Feautrier difference equations are given in \citet{Bommier-19} in the case of polarized radiation and density matrix formalism. However, in the multilevel case, the solution has to remain iterative, and was found to be overly time- and memory consuming to be applied to the present computation. In simpler trials, ignoring PRD or hyperfine structure, the convergence was found not to be significantly improved with respect to the convergence of the lambda-iteration method we applied here, which includes the short characteristics method renewed by \citet{Ibgui-etal-13} for radiative transfer equation integration.

\subsection{Convergence results}

We study the outgoing radiation polarization degree $Q/I$, which is the result of our calculation. The behavior of this quantity as a function of the iteration step is represented in Fig. \ref{convergence}, in the line center and in the line wings (at 5888.4 \AA\ for \ion{Na}{i} D$_2$ and 5894.9 \AA\ for \ion{Na}{i} D$_1$). When polarization quickly converges in the line center, this is not the case in the line wings. 

In order to demonstrate that convergence exists in line wings, we compare the behavior of our series with the behavior of the Riemann series:
\begin{equation}
% MathType!MTEF!2!1!+-
% faaahqart1ev3aaaKnaaaaWenf2ys9wBH5garuavP1wzZbItLDhis9
% wBH5garmWu51MyVXgaruWqVvNCPvMCaebbnrfifHhDYfgasaacH8sr
% ps0lbbf9q8WrFfeuY-ribbf9v8qqaqFr0xc9pk0xbba9q8WqFfea0-
% yr0RYxir-Jbba9q8aq0-yq-He9q8qqQ8frFve9Fve9Ff0dc9Gqpi0d
% meaabaqaciGacaGaaeqabaWaaeaaeaaakeaacaWGMbGaeyypa0Zaaa
% buaeaadaWcaaqaaiaaigdaaeaacaWGUbWaaWbaaSqabeaacqaHXoqy
% aaaaaaqaaiaad6gaaeqaniabggHiLdaaaa!3A05!
f = \sum\limits_n {\frac{1}{{{n^\alpha }}}} \ \ .
\end{equation}
When $\alpha = 1$, this is the harmonic series, which does not converge. However, when $\alpha > 1$ strictly, this is a Riemann series, which is absolutely convergent. 

We therefore compare the step increment of our series to the inverse of the step number $1/N$ in order to determine whether or not $\alpha$ is larger than unity. The top panel of Fig. \ref{riemann} shows the ratio of this increment $\Delta (Q/I)$ to $1/N$, as a function of the iteration step number $N$. For sufficiently large step numbers ($N>60$), $\Delta (Q/I)$ can be seen to decrease more quickly than $1/N$, which means that $\Delta (Q/I)$ behaves as $1/N^\alpha$ with $\alpha > 1$, meaning that our series is of the Riemann type, and is therefore absolutely convergent even in the line wings. 

In order to evaluate $\alpha$ in the case of our series, in the bottom panel of Fig. \ref{riemann} we plot the logarithm of the ratio, and we see that this logarithm decreases linearly  for both wings, which determines $\alpha = 1.00058$ for the D$_1$ wing and $\alpha = 1.0012$ for the D$_2$ wing, which are both larger than unity. As can be seen in Fig. \ref{riemann}, the slope of the linear interpolation is found to be significantly larger than the numerical noise. The series is therefore absolutely convergent. However, the behavior of the series and of its convergence remains unverified for the next iteration steps.

Figure \ref{converge} shows that for the 166 iteration steps we performed, the outgoing polarization $Q/I$ is not far from convergence in the line wings, but is not yet completely reached. These computations are rather time consuming, and we spent 26 Mh of computations in the TURING machine at IDRIS\footnote{Institut du D\'eveloppement et des Ressources en Informatique Scientifique, Orsay, France} to obtain these 166 iteration steps; we therefore estimate the above convergence to be sufficient. Convergence is however not fully reached in the line wings within the allocated computing time. Development of efficient acceleration methods would be desirable for future work.

The two terms of the $Q/I$ ratio, namely the Stokes parameters $Q$ and $I$, each nevertheless converge much less quickly, in such a way that they are not at all converged within the 166 iteration steps. However, the behavior of the ratio $Q/I$ appears to be different, because some simplifications certainly occur when forming the ratio. 

\section{Concluding discussion}
\label{discussion}

Although the calculation seems reasonably converged in the line wings, the agreement between the observed and computed profiles remains only qualitative. As described below, we investigated the different possible causes of this difference.

We questioned the accuracy of collision rate calculations, in particular those with hydrogen neutral atoms that determine the coherent scattering weight in PRD, and therefore in the far wings. We artificially varied these rates, but the \ion{Na}{i} D$_2$ wing profile was not improved with respect to observations. 

We questioned the model electron density, whose determination is very indirect from observations, and in light of the recent investigation by \citet{Bommier-20}, who suggests that the electron density at the solar surface would be much higher than in present models, and similar to the neutral hydrogen atom density. We artificially varied the model electron density, but again the \ion{Na}{i} D$_2$ wing profile was not improved with respect to observations. 

We questioned the accuracy of our model of Rayleigh scattering on neutral hydrogen atoms, which is responsible for the continuum below spectral lines, which is of more relative importance in their wings. Our model is taken from the MALIP code for computing the Stokes parameter profiles of magnetoactive Fraunhofer lines by \citet{Landi-76}, and we did not try to modify it.

We questioned the very rough HOLMUL approximation \citep{Holweger-Muller-74} we applied to our atmosphere model. This approximation consists in not considering atmosphere higher than the temperature minimum. The temperature reversal there may be responsible for a central bump in the line profile, following a well-known result for a two-level atom in LTE. As our model considers only the lower and upper levels of the  \ion{Na}{i} D lines (with all their sublevels and coherences), this is probably the reason why this bump appeared in our calculations, as visible in Fig. 12 of \citet{Bommier-16b}, which is not the case in observations. One solution for this problem could lie in considering several upper levels of \ion{Na}{i}, as done by \citet{Bruls-etal-92}. \citet{Leenaarts-etal-10} suggested modeling the contribution of several higher upper levels by grouping them into a single artificial upper level. We tried to implement this suggestion but it was unsuccessful because we treated the radiative transfer only in the \ion{Na}{i} D lines to remain within computing time restrictions.

We question the 1D plane parallel character of the usual atmosphere model above the temperature minimum region. In hydrogen H$\alpha$ images, this part of the atmosphere contrarily seems highly inhomogeneous and structured into fibrils. Also inspired by the first ionization potential (FIP) effect observed higher in the corona, we tried an increase of the atmosphere model electron density in this part of the atmopshere, as if the electrons were grouped into inhomogeneous structures along the line of sight above the temperature minimum region. This was unsuccessful.

We finally note that the \ion{Na}{i} D line intensity profiles are broader in our computation results than in all the observations. The central peak of the \ion{Na}{i} D$_2$ line polarization is accordingly broader. This may be the reason why the \ion{Na}{i} D$_2$ wing bump is much less visible in our results than in observations. In relation, we question the fact that we neglected the inhomogeneous upper part of the atmosphere. In other words, we assign this discrepancy to the inadequacy of the plane parallel atmosphere model to describe the chromosphere, where the \ion{Na}{i} D lines are formed. We think that this is the main limitation of our computation. As all integrals over frequencies and velocities, and their directions, are numerical, the calculation could probably be adapted to more realistic and higher dimensional model atmospheres. 

Nevertheless, our approach is the first self-consistent treatment of redistribution in NLTE conditions. The lower level atomic polarization is fully accounted for, although found to be negligible, unlike what was assumed by \citet{Landi-98}; it is destroyed by collisions that are fully accounted for. However, in the case of a line with hyperfine structure, as in the \ion{Na}{i} D lines, the calculation is time consuming.

This work was motivated by the existence of a net linear polarization in \ion{Na}{i} D$_1$, as observed by \citet[Fig. 2]{Stenflo-Keller-97} and \citet[Fig. 1]{Stenflo-etal-00a}, as well as by \citet[Fig. 1]{Stenflo-etal-00b}. On the contrary, \citet[Figs. 2, 3]{Bommier-Molodij-02}, \citet[Figs. 2, 3]{TrujilloB-etal-01}, and \citet{Gandorfer-00} did not observe any such net linear polarization in \ion{Na}{i} D$_1$. As expected from the \ion{Na}{i} D$_1$ upper level kinetic momentum quantum number $J=1/2$, our computation does not reveal any net linear polarization in \ion{Na}{i} D$_1$. We note that \citet{Bommier-Molodij-02} and \citet{TrujilloB-etal-01} both observed with the TH\'EMIS\footnote{T\'elescope H\'eliographique pour l'\'Etude du Magn\'etisme est des Instabilit\'es Solaires} telescope, which was polarization-free at that time, and that both works did not apply the same data-reduction code. \citet{Gandorfer-00} observed at IRSOL\footnote{Istituto Ricerche Solari Locarno}, where the telescope is polarization-free around equinox, whereas \citet{Stenflo-Keller-97}, \citet{Stenflo-etal-00a}, and \citet{Stenflo-etal-00b} observed with the Z\"urich Imaging Polarimeter ZIMPOL mounted on the McMath-Pierce telescope at Kitt Peak, which has large and varying instrumental polarization \citep[p. 782]{Stenflo-etal-00b}. Instrumental polarization correction was installed with a tilting glass plate. The compensation level is about 0.1\% \citep[p. 782]{Stenflo-etal-00b}, which is also the order of magnitude of the \ion{Na}{i} D$_1$ linear polarization. Therefore, the compensation may be insufficient given the low linear polarization of the \ion{Na}{i} D lines, and the net linear polarization observed in \ion{Na}{i} D$_1$ with this instrument only, lacking confirmation by theoretical calculations, could be an artefact of instrumental polarization. 

In addition, as already stated at the end of Sect. \ref{results}, the linear polarization observed by \citet[Fig. 2]{Stenflo-Keller-97} and \citet[Fig. 1]{Stenflo-etal-00a} in the \ion{Na}{i} D$_2$ line center is also higher by 0.1\% than the present calculation result. This cannot be assigned to a magnetic field effect not accounted for in our calculations because the Hanle effect acts as a depolarizing effect in the second solar spectrum. The difference in polarization degree between our calculation result and these observations is about 0.1\%, which is comparable to the instrumental compensation level. The linear polarization in \ion{Na}{i} D$_1$ similarly shows a narrow peak of 0.1\% linear polarization at line center, which is not present in the calculation result. The 0.1\% excess polarization observed in both line centers could therefore be an artefact of instrumental polarization.

\begin{acknowledgements}
For set up, this work was granted access to the HPC resources of MesoPSL financed
by the Region Ile de France and the project Equip@Meso (reference
ANR-10-EQPX-29-01) of the programme Investissements d'Avenir supervised
by the Agence Nationale pour la Recherche (France). For exploitation, this work was granted access to the HPC resources of IDRIS under the allocation 2017-A0020407205 made by GENCI (France).
\end{acknowledgements}

\bibliographystyle{aa}

\end{document}